\documentclass[aps,pre,reprint,amsmath,amsfonts,superscriptaddress]{revtex4-1}
\usepackage[utf8]{inputenc}
\usepackage[english]{babel}
\usepackage{times}
\usepackage[colorlinks,citecolor=blue]{hyperref}
\usepackage{graphicx}

\providecommand{\abs}[1]{\ensuremath{\lvert #1 \rvert}}
\providecommand{\avk}{\ensuremath{\langle k \rangle}}
\providecommand{\avqo}{\ensuremath{\langle q_{\rm o} \rangle}}
\providecommand{\kap}{\ensuremath{\kappa}}
\providecommand{\kr}[2]{\ensuremath{\delta_{#1 #2}}}
\providecommand{\mto}{\ensuremath{\mspace{-2mu}{\rightarrow}\mspace{-2mu}}}
\providecommand{\mfrom}{\ensuremath{\mspace{-2mu}{\leftarrow}\mspace{-2mu}}}
\providecommand{\ld}[2]{\ensuremath{#1\mto #2}}
\providecommand{\lab}{\mathrm{\ld{A}{B}}}
\providecommand{\lba}{\mathrm{\ld{B}{A}}}
\providecommand{\laa}{\mathrm{\ld{A}{A}}}
\providecommand{\lbb}{\mathrm{\ld{B}{B}}}

\providecommand{\fref}[1]{Fig.~\ref{#1}}

\addto\captionsenglish{}

\begin{document}
\title{Early fragmentation in the adaptive voter model on directed networks}

\author{Gerd Zschaler}
 \email{zschaler@pks.mpg.de}
\author{Gesa A.\ B\"ohme}
\author{Michael Sei\ss{}inger}
 \affiliation{Max-Planck-Institut f\"ur Physik komplexer Systeme, N\"othnitzer Str. 38, 01187 Dresden, Germany}
\author{Cristi\'an Huepe}
 \affiliation{614 N. Paulina Street, Chicago Illinois 60622-6062, USA}
\author{Thilo Gross}
 \affiliation{Department of Engineering Mathematics, Merchant Venturers Building, University of Bristol, Woodland Road, Clifton, Bristol BS81TR, United Kingdom}

\begin{abstract}
We consider voter dynamics on a directed adaptive network with fixed out-degree distribution. A transition between an active phase and a fragmented phase is observed. This transition is similar to the undirected case if the networks are sufficiently dense and have a narrow out-degree distribution. However, if a significant number of nodes with low out-degree is present, then fragmentation can occur even far below the estimated critical point due to the formation of self-stabilizing structures that nucleate fragmentation. This process may be relevant for fragmentation in current political opinion formation processes.
\end{abstract}

\maketitle

\section{Introduction}
The defining feature of complex systems is the emergence of collective phenomena from the interaction of many parts \cite{Anderson1972}. A vivid example is provided by social (swarm) intelligence \cite{Couzin2007}. Crowds of humans, shoals of fish, and even swarms of insects are known to solve problems more efficiently than any individual on its own \cite{Sumpter2008, Huepe2011}. However, only in humans social intelligence is also used on a higher level. Among all species, only we have evolved the ability to discuss (and debate) future problems and opportunities, and formulate long-term policy.

Recent media reports have pointed out that a central function of the political opinion formation process, the debunking of counter-factual opinions, may be starting to fail (see \cite{Maddow2010} for an example from U.S.\ politics).

In the political discourse our opinions are exposed to close scrutiny and criticism. Well-founded criticism may cause us to change counter-factual beliefs and thus promote rational decision making. However, because of the stresses involved, humans tend to favor discussing with others who share similar beliefs.

With the increasing diversity of offline and online media \cite{Webb2006, Flaounas2010} and new media technologies \cite{Hensinger2010}, it is becoming easier to avoid opposing opinions altogether. In particular, the Internet enables people not only to access but also to publish information easily. One of the best examples is perhaps the micro-blogging service Twitter, which currently is approaching $10^9$ user posts per week. Among this flood of information, it is easy to find sources supporting almost every conceivable opinion, while avoiding contradicting evidence.

It thus seems likely that situations develop where a given subset of the society (and the media by which it is represented) pay attention only to information sources with the same belief system, thus reinforcing and perpetuating myths that are never confronted with opposing views. In this light, one may ask whether we are heading for a society that is fractionated into groups adhering to internally consistent but mutually exclusive belief systems.

The question when fragmentation in opinion formation processes can occur on a social scale is beyond the scope of classical social research, because it focuses on an emergent phenomenon that may require new conceptual approaches from physics.
A minimal yet paradigmatic physical model of opinion formation is the voter model \cite{Holley1975, Liggett1999}. It describes a network of nodes representing agents, and links representing the social contacts among them. The agents hold one of two possible opinions, which they can change by adopting the opinion of their topological neighbors. Due to its similarity to interacting spin models, the voter model has attracted considerable attention in the physics literature and has been studied using different interaction geometries, such as regular lattices or heterogeneous networks \cite{Sood2005, Castellano2005, Sood2008, Pugliese2009}.

An important extension of the voter model is achieved by including homophily, the agents' propensity to discard links to opposing neighbors and establish new links to agents holding the same opinion \cite{Ehrhardt2006}. Thus, the interaction network is not static but co-evolves with the agent dynamics, as the agents rewire their links depending on their opinions. Such networks, in which the node and link dynamics co-evolve, are called adaptive networks \cite{Gross2008, Gross2009, ANWiki}.

The long-term dynamics of the adaptive voter model can reach one of two absorbing states: a \emph{consensus state}, in which all agents hold the same opinion, or a \emph{fragmented state}, in which the network breaks into at least two components, which are internally in consensus. The adaptive voter model therefore provides a simple framework in which the fragmentation of opinion formation processes can be studied \cite{Holme2006, Vazquez2008, Nardini2008, Kimura2008, Kozma2008, *Kozma2008a}. In this model, fragmentation occurs in a phase transition that has been identified as a generic absorbing transition \cite{Vazquez2008}. It can be computed analytically using a recently proposed motif expansion approach \cite{Boehme2011}.

The adaptive voter model has been studied so far on undirected networks. In the context of opinion formation, however, the underlying interactions are often asymmetric. It is therefore reasonable to encode ``who pays attention to whom'' as directed links in the interaction network. Directed links were considered in previous studies on static networks \cite{Sanchez2002, Park2006, Jiang2008, Serrano2009, Masuda2009}, but directed adaptive networks have only been investigated in a generic threshold model for Boolean networks \cite{Lambiotte2011}.

In this paper, we investigate voter dynamics in a directed adaptive network, in which both the opinion dynamics and the topological change are influenced by the directionality of the interactions. The agents can avoid disagreeing with neighbors by rewiring their outgoing links, whereas they cannot affect their incoming links. The overall network topology thus changes adaptively, while the out-degree of each agent, i.e., its number of outgoing links remains unchanged. This enables us to study the influence of different realistic out-degree distributions.

For sufficiently dense Poissonian out-degree distributions, fragmentation occurs if the rewiring of network connections exceeds a critical rate, which is consistent with previous results on undirected networks \cite{Holme2006, Vazquez2008, Nardini2008, Kimura2008, Kozma2008, *Kozma2008a}. However, for scale-free out-degree distributions and Poissonian distributions with small mean degree, we find that fragmentation can already be observed at much lower rewiring rates than in undirected networks. We show that this behavior is due to the nodes of low out-degree, which can form self-stabilizing topological structures that nucleate fragmentation.

\section{Model}
We consider a network of $N$ nodes representing agents and $K$ directed links representing social contacts. Each node $i$ holds a binary opinion $\sigma_i \in \{\mathrm{A, B}\}$. The direction of links indicates the flow of attention between the agents. In other words, in our notation we draw links in the direction that one would draw the ``follows''-links on Twitter.

We initialize the network as a random directed graph with randomly assigned equiprobable strategies. The node states and the network topology are then left to evolve according to the following rules: In sequential updates, a link $i \mto j$ is picked at random from the network \footnote{In this work, we use the link update scheme. We have checked that the results reported here do not change qualitatively when using node update, i.e., selecting a random node and then one of its outgoing neighbors.}. If $\sigma_i = \sigma_j$, the link is said to be \emph{inert} and nothing happens. Otherwise, the link is said to be \emph{active}, and it is either rewired (probability $p$), or an opinion update takes place (probability $1-p$). In the former case, the node $i$ cuts the link and reconnects to a random node $k$ with $\sigma_k=\sigma_i$. In the latter case, node $i$ switches its opinion $\sigma_i$ to $\sigma_j$. We note that the rewiring of links changes only the in-degree distribution, whereas the out-degree distribution and the average degree $\avk = K/N$ of the network remain fixed.

In contrast to previous studies of the voter model on static directed networks, we do not need to restrict our model to networks consisting of a single strongly connected component \cite{Serrano2009, Sanchez2002, Dorogovtsev2001}, because the network's component structure is affected by the ongoing rewiring of links, which continuously forms and re-routes paths between different strongly connected components.

Below, we study the proposed model in terms of the density $n$ of $\mathrm{A}$-nodes (corresponding to agents holding opinion A) and the per-capita densities $f$ and $h$ of active links, $f\equiv [\lab]$ and $h \equiv [\lba]$. Following \cite{Vazquez2008, Nardini2008} we characterize the state of the network by the magnetization $m = 1-2n$ and the active link density $\rho = (f+h)/\avk$.

In network simulations of the directed adaptive voter model one observes qualitatively different types of trajectories: First at sufficiently high rewiring rates the network rapidly approaches a fragmented state ($\abs{m} > 0$, $\rho=0$), in which the network breaks into at least two components, which are internally in consensus. Second, for sufficiently low rewiring rates the network initially approaches a stationary active state ($|m| \ll 1 $, $\rho > 0$), in which the opinions and the topology change continually. Because such active states form a parabola in the $\rho$-$m$-plane, the system can drift randomly along the parabola until an absorbing consensus state ($m=\pm1$, $\rho=0$) is reached. These dynamics are closely reminiscent of the adaptive voter model on undirected networks \cite{Vazquez2008,Kimura2008}.

\begin{figure}
 \includegraphics[width=0.48\textwidth]{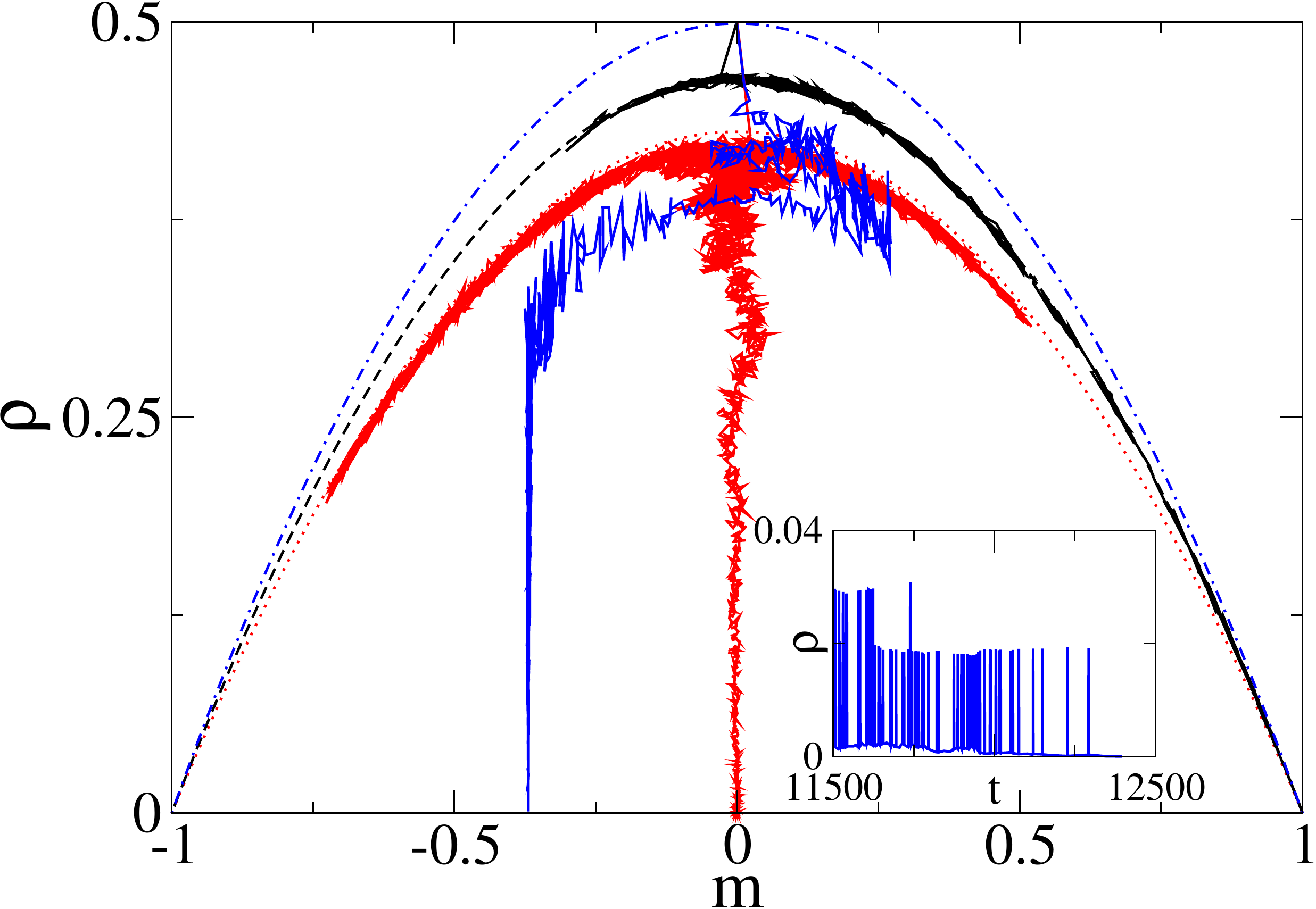}
 \caption{\label{fig:traj-lu}(Color online) Typical trajectories from network simulations. The state of the network is characterized by the density of active links $\rho$ and the magnetization $m$. The trajectories shown correspond to networks with Poissonian out-degree distributions with $\avk=4$ (red/light gray), $\avk=8$ (black), and an out-degree distribution following $P_{\rm out}(k)\propto k^{-2}$ (blue/dark gray).
The trajectories initially drift along a parabola of active states (dotted red, dashed black, and dash-dotted blue lines, denoting analytical results from Eq.~\eqref{eq:ss1a}).
However, only the black trajectory reaches a consensus state, whereas the others eventually collapse to a fragmented state. The inset shows a time series of $\rho$ from the scale-free network shortly before fragmentation. The parameters are $N=10^4$ and $p=0.1$.}
\end{figure}

In addition to the trajectories described above, the directed model can show a third type of behavior not observed in the undirected case. Here, the systems drifts along the parabola of active states for some time and then collapses slowly to the fragmented state (Fig.~\ref{fig:traj-lu}). This can lead to fragmentation significantly below the critical rewiring rate found in undirected networks. The delayed fragmentation after the drift along the parabola of active states suggests that in the active state the network undergoes some slow reorganization that eventually leads to the destabilization of the active states. In the following we investigate the nature of this reorganization and its implications for network fragmentation.

\section{Analytical approach}
The main goal of this paper is exploring the impact of directionality of attention on the opinion formation process. For this purpose we compare the dynamics of the directed adaptive voter model to well known results for the undirected adaptive voter model. In the following we refer to these two models simply as the \emph{directed model} and the \emph{undirected model}, respectively. A direct comparison of different models is often difficult and may lead to misleading results. We therefore compare simulation results of the directed model to two analytical approximations that are known to capture the dynamics of the undirected model in different limits. In this comparison an agreement between analytical and numerical results indicates that the assumptions made for the undirected model are still valid in the directed model, whereas a disagreement points to new physics in the directed model that is not observed in the undirected model.

\subsection{Moment expansion}
The undirected model was studied extensively by moment expansions \cite{Gross2006, Vazquez2008, Kimura2008}. Following previous works, we derive differential equations for the time evolution of so-called network moments, namely the densities $n$, $f$, and $h$ defined above.
The change in the density of A-nodes, $n$, is given by the balance between opinion adoption in $\lba$ and $\lab$ links,
\begin{equation}
 \label{eq:nfh-ode0}
 \dot{n}=(1-p)\left( h - f \right).
\end{equation}
The density of $\lab$ links, $f$, changes according to
\begin{multline}
\label{eq:mc2}
 \dot{f} = -pf + (1-p) \bigl\{ [\laa\mto\mathrm{B}] - 2[\mathrm{B}\mfrom\lab] \\
  + [\mathrm{B}\mfrom\lba] - [\lab\mto\mathrm{A}] - f \bigr\},
\end{multline}
where $[\ld{X}{Y}\mto Z]$ and $[\mathrm{X}\mfrom\ld{Y}{Z}]$ denote the per-capita densities of triplets in the network ($X,Y,Z\in\{\mathrm{A},\mathrm{B}\}$). In Eq.~\eqref{eq:mc2}, the first term corresponds to the gain in $f$ due to rewiring, whereas the remaining terms correspond to gains and losses due to opinion adoption. In an opinion adoption event, a node copies a neighbor's opinion via one of its outgoing active links, transforming it into an inert link. This also affects all other links connected to the focal node, transforming active links into inert ones and vice versa. The resulting indirect change in the density of active links is accounted for by the triplet variables.

The time evolution of the density of $\lba$ links, $h$, is determined by the analogous equation
\begin{multline}
\label{eq:mc3}
 \dot{h} = -ph + (1-p) \bigl\{ [\lbb\mto\mathrm{A}] - 2[\mathrm{A}\mfrom\lba] \\
  + [\mathrm{A}\mfrom\lab] - [\lba\mto\mathrm{B}] - h \bigr\}.
\end{multline}

Equations~\eqref{eq:nfh-ode0}--\eqref{eq:mc3} do not constitute a closed ordinary differential equation (ODE) system, as they involve the triplet moments $[\ld{X}{Y}\mto Z]$ and $[X\mfrom\ld{Y}{Z}]$. In principle, the equation system could be complemented by similar equations for the triplet moments. These would, however, depend on higher moments, such as four-node motifs. An appropriate truncation of the expansion is thus necessary in order to obtain a closed system of equations. This approximation is referred to as moment closure \cite{Keeling2005, Gross2006}.

In the following, we express the triplet densities in terms of node and link densities using the pair approximation \cite{Keeling2005, Gross2006, Zschaler2010, Demirel2011a},
\begin{align}\label{eq:pa-asymm}
 [\ld{X}{Y}\mto Z]  &\approx [\ld{X}{Y}]\avk\frac{[\ld{Y}{Z}]}{\avk[Y]}, \\
 \label{eq:pa-symm}
 \mu_{XZ}[X\mfrom\ld{Y}{Z}] &\approx [\ld{Y}{X}]\avqo\frac{[\ld{Y}{Z}]}{\avk[Y]},
\end{align}
where $\mu_{XZ} = 1+\kr{X}{Z}$ accounts for the double-counting of symmetric triplets. In the equations, the numbers of triplets are approximated as the number of $\ld{X}{Y}$ or $\ld{Y}{X}$ links times the average number of $\ld{Y}{Z}$ links connected to a $Y$-node. Here, we assume that the probability of finding an outgoing $Z$-neighbor of a $Y$-node is independent of the presence of an $X$-neighbor of the $Y$-node. Thus the probability of finding a given triplet depends on the global link density $[\ld{Y}{Z}]/\avk [Y]$. In \eqref{eq:pa-asymm}, each of the $\avk$ outgoing links of the $Y$-node is a $\ld{Y}{Z}$ link with this probability. In \eqref{eq:pa-symm}, in contrast, the $Y$-node has already been selected by following one of its outgoing links. In this case, each of its remaining $\avqo$ outgoing links is a $\ld{Y}{Z}$ link with this probability. The quantity $\avqo$ is the mean excess out-degree in the network, which can be computed from the out-degree distribution \cite{Newman2003}.

As a further simplification we assume that the mean degree of both node types is equal, so that $[\laa] = \avk n - f$ and $[\lbb] = \avk(1-n) - h$ \cite{Vazquez2008}. We then obtain a closed set of ODEs,
\begin{equation}
\label{eq:nfh-ode1}
 \dot{n}=(1-p)\left( h - f \right),
\end{equation}
\begin{multline}
\label{eq:nfh-ode2}
\dot f = -pf +(1-p)\biggl\{ \frac{(\avk n-f)f}{n} -f \\
-\frac{fh}{1-n} +\kap \frac{(\avk (1-n)-h)h}{1-n} -\kap \frac{f^2}{n} \biggr\},
\end{multline}
\begin{multline}
\label{eq:nfh-ode3}
\dot h = -ph +(1-p)\biggl\{ \frac{(\avk (1-n)-h)h}{1-n} -h \\
-\frac{fh}{n} +\kap \frac{(\avk n-f)f}{n} - \kap \frac{h^2}{1-n} \biggr\},
\end{multline}
where $\kap = \avqo/\avk$.

This system of differential equations has a trivial solution,
\begin{equation}
\label{eq:ss1b}
 h^*=f^*=0,
\end{equation}
corresponding to the absorbing states, in which no active links are left. These states can be either fragmented states ($0<n^*<1$) or consensus states ($n^*=0$ or $1$).

Additionally, there is a continuum of non-absorbing, active stationary states,
\begin{equation}
\label{eq:ss1a}
 h^*=f^*=n^*(1-n^*)\left(\avk - \frac{1}{(1+\kap)(1-p)}\right),
\end{equation}
in which neither consensus nor fragmentation is achieved. These states form the parabola in the $\rho$-$m$-plane discussed above (\fref{fig:traj-lu}).

In the undirected model the moment expansion is known to yield good results when $p$ is far from the
fragmentation point, but to become less precise for $p$ close to fragmentation. To estimate the fragmentation point we
therefore resort to a motif expansion described below. It is nevertheless instructive to consider the critical rewiring rate
\begin{equation}\label{eq:pcrit}
\tilde p_c = 1- \left[\avk (\kap +1)\right]^{-1},
\end{equation}
which is computed by a linear stability analysis of the active states. For $p<\tilde p_c$ the active states are stable, whereas the absorbing states are stable for $p>\tilde p_c$. This estimate of the critical point closely resembles
the analogous result for the adaptive voter model using link update on undirected networks \cite{Demirel2011a}. The main difference is that in the undirected case, the parameter $\kap$ is unknown and therefore usually set to unity. This approximation is well justified, because in the undirected model, the ongoing rewiring leads to an approximately Poissonian degree distribution for which $\kap=1$ \cite{Newman2003}. In contrast, in the directed model, the outgoing degree distribution remains fixed, and $\kappa$ has to be considered explicitly. In the results presented here, we use values of $\kap$ that are explicitly measured in realizations of the respective out-degree distributions.

\subsection{Motif expansion}
\begin{figure}
 \includegraphics[width=0.48\textwidth]{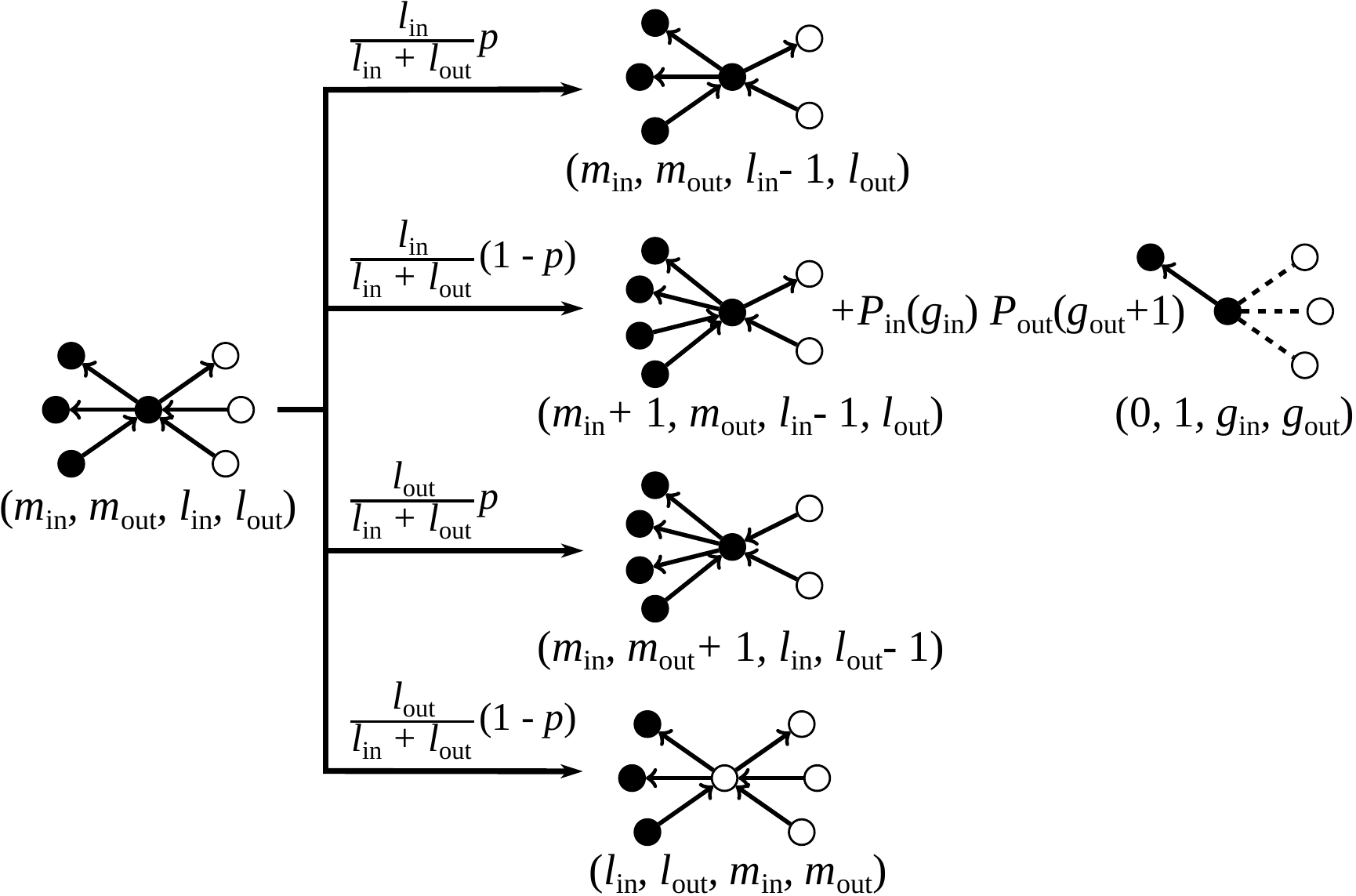}
 \caption{\label{fig:perc-appr-scheme}Transitions of a general active motif $(m_{\rm in}, m_{\rm out}, l_{\rm in}, l_{\rm out})$. The transition probabilities follow from the link update rule of the directed adaptive voter model. New active motifs are created only when an opinion update occurs on an incoming active link (second transition). In this case the number of incoming and outgoing active links (dashed) of the new motif is estimated based on the in-degree distribution $P_{\rm in}$ and out-degree distribution $P_{\rm out}$ of the underlying network.}
\end{figure}
In the undirected model a precise estimate of the transition point can be obtained by a \emph{motif expansion} proposed in Ref.~\cite{Boehme2011}. In this expansion we consider an almost fragmented network, consisting of two almost isolated components that are internally in consensus, but are still connected by a low density of ``active motifs''. In the directed network the active motifs are characterized by their numbers of inert incoming and outgoing links ($m_{\rm in}, m_{\rm out}$) and active incoming and outgoing links ($l_{\rm in}, l_{\rm out}$). Following \cite{Boehme2011}, we derive a set of balance equations capturing the effect of all possible update processes on the densities of active motifs,
\begin{align}
 &\dot\rho(m_{\rm in}, m_{\rm out},l_{\rm in}, l_{\rm out}) = -\rho(m_{\rm in}, m_{\rm out},l_{\rm in}, l_{\rm out}) \notag\\
 &\quad+\frac{l_{\rm in}+1}{l_{\rm in}+1+l_{\rm out}} p \rho(m_{\rm in}, m_{\rm out},l_{\rm in}+1, l_{\rm out}) \notag\\
 &\quad+\frac{l_{\rm in}+1}{l_{\rm in}+1+l_{\rm out}} (1-p)\rho(m_{\rm in}-1, m_{\rm out},l_{\rm in}+1, l_{\rm out}) \notag\\
 &\quad+\frac{l_{\rm out}+1}{l_{\rm in}+l_{\rm out}+1} p \rho(m_{\rm in}, m_{\rm out}-1,l_{\rm in}, l_{\rm out}+1) \notag\\
 &\quad+\frac{m_{\rm out}}{m_{\rm in}+m_{\rm out}} (1-p) \rho(l_{\rm in}, l_{\rm out},m_{\rm in}, m_{\rm out})
\end{align}
for $m_{\rm in}>0$ and $m_{\rm out}>1$, and
\begin{multline}
 \dot\rho(0,1,l_{\rm in}, l_{\rm out}) =-\rho(0,1,l_{\rm in}, l_{\rm out}) \notag \\
 +(1-p) P_{\rm in}(l_{\rm in}) P_{\rm out}(l_{\rm out}+1) \notag \\
 \times\sum\frac{n_{\rm in}}{n_{\rm in}+n_{\rm out}}\,\rho(m_{\rm in}, m_{\rm out},n_{\rm in}, n_{\rm out})
\end{multline}
for $m_{\rm in}=0$ and $m_{\rm out}=1$. The summation runs over all active motifs $(m_{\rm in}, m_{\rm out},n_{\rm in}, n_{\rm out})$ up to maximum in- and out-degrees, i.e., over all possible 4-tuples with
 $m_{\rm in}+n_{\rm in}\le \hat k_{\rm in}$ and $m_{\rm out}+n_{\rm out}\le \hat k_{\rm out}$, where $\hat k_{\rm in}$ and $\hat k_{\rm out}$ denote the cut-offs. Note that the dimension of the transition matrix grows with the cut-off faster than ${\hat k}^3$, which has a significant effect on the computation time.

A schematic representation of the transition probabilities is shown in \fref{fig:perc-appr-scheme}. We account for heterogeneous in- and out-degree distributions ($P_{\rm in},P_{\rm out}$), but assume that the in- and out-degree of a node are uncorrelated. In the balance equations, the fragmented state is obtained as the stationary solution containing zero active motifs. The critical rewiring rate $p_c$ is then extracted from the linear stability analysis of this state as the rewiring rate at which the fragmented state becomes stable.

\section{Numerical exploration of early fragmentation}
In the following we compare the estimated fragmentation points, obtained from the approximations above, with results from agent-based simulation of the networks. We first consider the case of a network with Poissonian out-degree distribution with $\avk=8$. As a second example we study a network with a scale-free out-degree distribution, in which the fragmentation occurs much earlier. We conjecture that this early fragmentation occurs due to the presence of a large number of nodes with low out-degree, which is then verified in a network with Poissonian degree distribution with $\avk=4$.

\subsection{Poissonian out-degree distribution with \texorpdfstring{$\boldsymbol{\avk=8}$}{<k> = 8}}
%
\begin{figure}
 \includegraphics[width=0.48\textwidth]{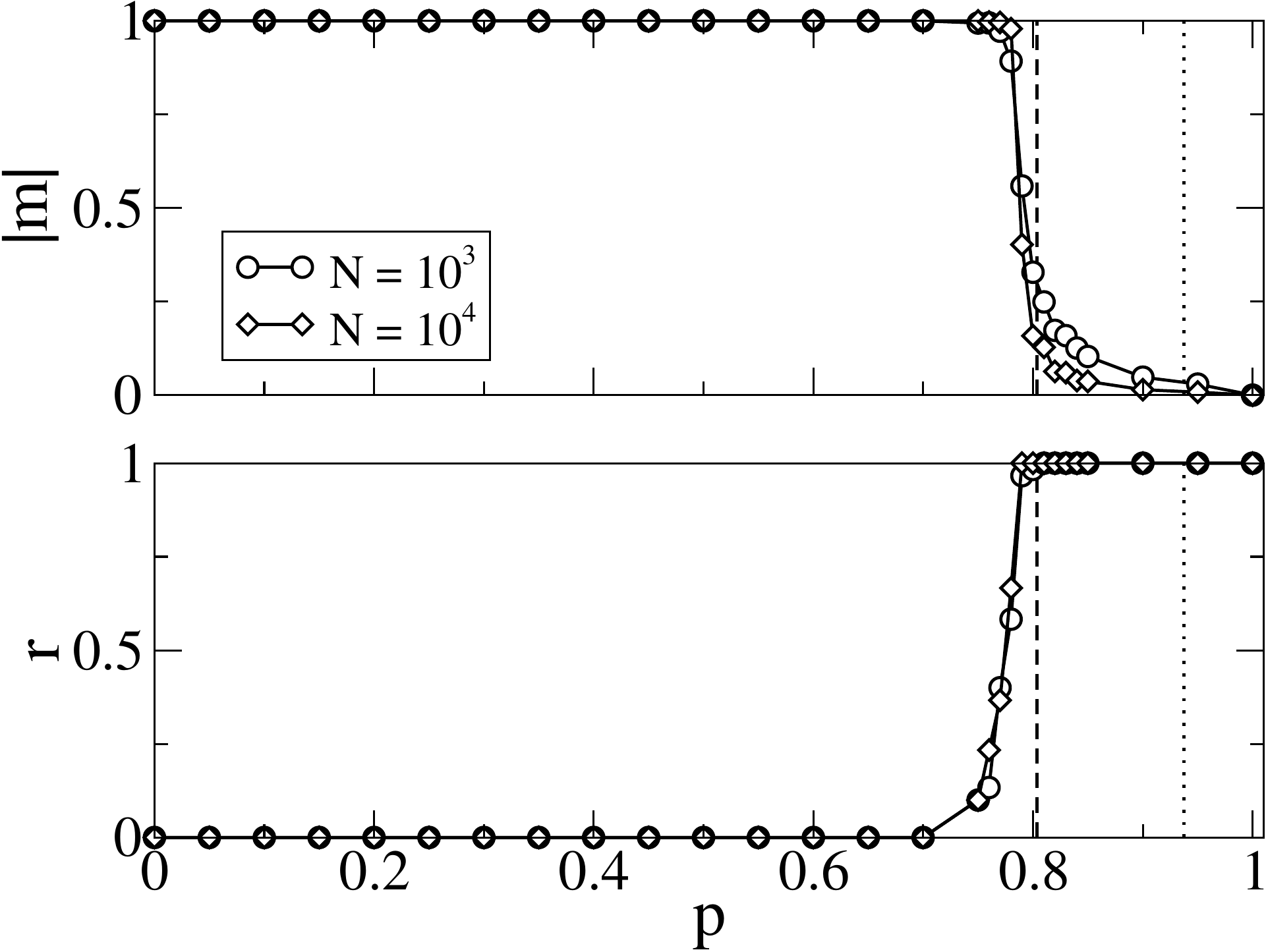}
 \caption{\label{fig:trans-mu8-lu}Fragmentation of a network with Poissonian out-degree distribution and $\avk=8$. Shown is the absolute value of the magnetization in the final frozen state (top) and the proportion $r$ of simulation runs that reaches a fragmented state before $t_{\rm max}=80\,000$ (bottom) as a function of the rewiring rate $p$. Each point is an average over 100 simulations. The critical point computed by the moment closure approximation (dotted line) overestimates the critical rewiring rate, whereas the motif expansion yields a better estimate of the true fragmentation point. For the motif-expansion a cut-off of $\hat k_{\rm in}=\hat k_{\rm out}=10$ was used. For a higher cut-off the estimated critical rewiring rate is expected to shift to slightly higher values.}
\end{figure}
We first consider networks with a Poissonian out-degree distribution, because this distribution closely matches the distribution observed in the undirected model \cite{Vazquez2008}. Starting from a random graph with both in- and out-degrees drawn from a Poisson distribution with mean $\avk$, we simulate the full network dynamics for systems of up to $N=10^4$ nodes until a frozen state is reached or a maximum simulation time $t_{\rm max}$ is exceeded. Time is measured in units of $1/K$, so that $K$ update events take place in one simulated time unit.

The results in \fref{fig:trans-mu8-lu} show a relatively sharp fragmentation transition at a critical rewiring rate $p_c\approx 0.79$. For $p<p_c$, the network reaches a state of global consensus, in which all nodes have the same state ($\abs{m}=1$). By contrast, for $p>p_c$, it separates into two disconnected components of approximately the same size, which hold opposing opinions but are internally in consensus. These results are strongly reminiscent of the undirected model \cite{Vazquez2008, Kimura2008, Boehme2011}.

The analogy between the directed and undirected model extends also to the analytical results. As in the undirected model the moment expansion overestimates the transition point, whereas the motif expansion yields a relatively precise estimate.

The study of the directed model in Poissonian networks with $\avk=8$ highlights the similarities between the directed and undirected networks and provides a basic test for our analytical approaches. For these networks the directed model exhibits the same dynamics as the undirected model and the analytical approaches capture the dynamics with similar precision as in the undirected case.

\subsection{Scale-free out-degree distribution}
\begin{figure}
\includegraphics[width=0.48\textwidth]{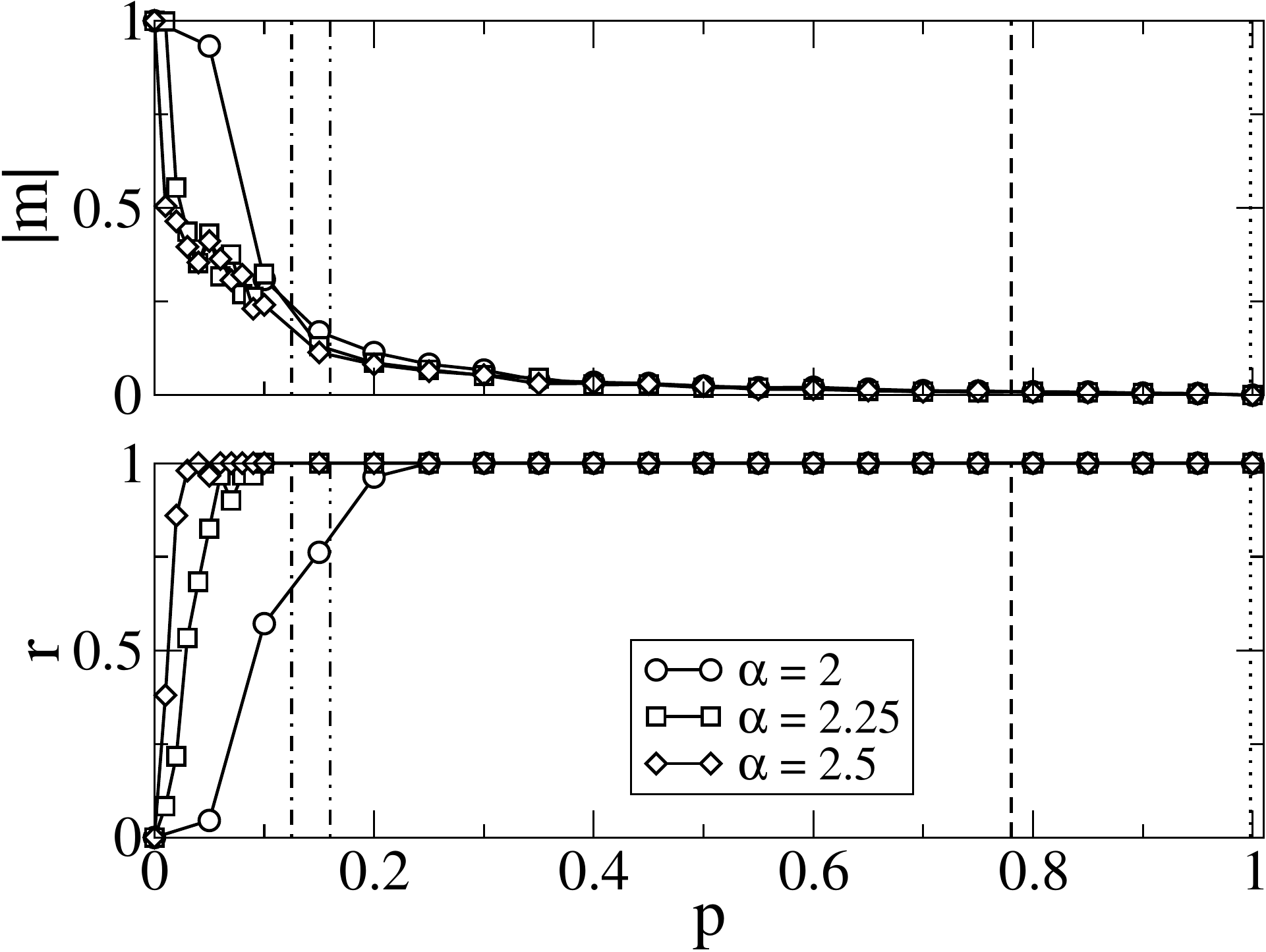}
 \caption{\label{fig:trans-a2-lu}Early fragmentation in scale-free networks. The plots are analogous to \fref{fig:trans-mu8-lu}, but describe networks with an out-degree distribution $P_{\rm out}(k)\propto k^{-\alpha}$. Fragmentation occurs far below the estimated transition points [for the motif expansion $\alpha=2$ (dashed line), $\alpha=2.25$ (dash--double-dotted line), $\alpha=2.5$ (dash-dotted line), and for the moment closure $\alpha=2$ (dotted line)] and extends over a wider range. The parameters are $N=10^4$, $\hat k_{\rm in}=\hat k_{\rm out} = 10$, and $\avk=5.5665$, $2.7042$, $1.876$ for $\alpha=2$, $2.25$, and $2.5$, respectively.}
\end{figure}
We now ask how the model behaves for more realistic out-degree distributions that cannot be realized in the previously studied undirected model. In the following, we consider power-law distributions of the form $P_{\rm out}(k)\propto k^{-\alpha}$, which capture the diversity that is observed in a wide variety of social applications \cite{Barabasi1999}. For generating networks with power-law distributed out-degrees and Poissonian in-degrees, we first draw an out-degree sequence of length $N$ from a power-law distribution. For each out-degree $k_i$ in this sequence, we then connect the outgoing links of node $i$ to $k_i$ random nodes in the network.  We explicitly avoid creating nodes without outgoing links as these would never change their state and act as ``zealots,'' trivially preventing the possibility of global consensus \cite{Mobilia2007}.

The results in \fref{fig:trans-a2-lu} show that in scale-free networks with $\alpha\in [2,3]$ the fragmentation occurs much earlier than in the Poissonian case. Moreover, the proportion $r$ of networks reaching fragmentation now increases gradually with increasing $p$. Considering individual simulation runs in detail one finds that the networks remain for some time in an active state before slowly approaching fragmentation---a behavior not observed in the undirected model or in the directed networks considered in the preceding section.

The observation that the networks spend some time in the active state before fragmenting indicates that these states are still feasible at least in the beginning of the simulation runs. The mechanism by which fragmentation is reached must therefore differ from the mechanism observed in the cases studied so far, where fragmentation occurs due to the destabilization of the parabola of active states in a transcritical bifurcation.

Notably both the moment and motif expansion seem not to capture the different mechanism for fragmentation because they significantly overestimate the fragmentation point. In the following we call fragmentation well below the estimated fragmentation point \emph{early fragmentation}. The main assumption used in both approximations is the absence of correlations between a node's in- and out-degree and between nearest neighbors. Their failure thus indicates the appearance of correlations that are absent or not substantial in the networks with Poissonian out-degree distribution.

\begin{figure}
 \includegraphics[width=0.48\textwidth]{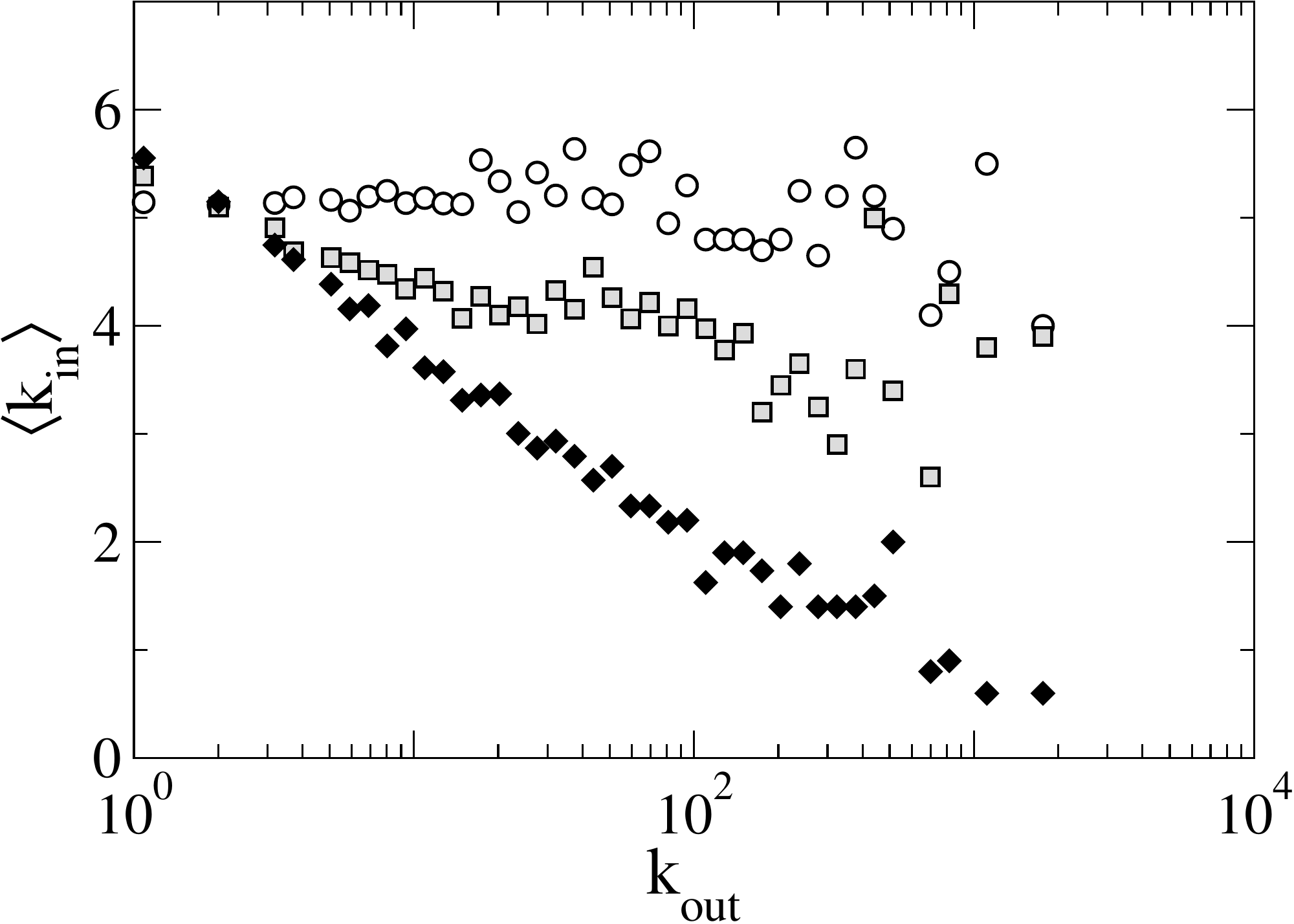}
 \caption{\label{fig:degrees} Average in-degree vs.\ out-degree at the initial ($t=0$, circles), intermediate ($t=20$, squares), and stabilized stage before full fragmentation ($t=500$, diamonds). Shown is the average in-degree of the network nodes vs.\ their logarithmically binned out-degree, averaged over 10 simulation runs. On average, the in-degree of nodes with large out-degree decreases with time. The parameters are: $N=5000$ and $P_{\rm out}(k) \propto k^{-2}$, $p=0.3$.}
\end{figure}

Network simulations suggest that early fragmentation is initiated by the formation of self-stabilizing structures among the agents. To understand the process leading to such structures, consider that the networks, even far from fragmentation, are partially ordered. On average the number of incoming and outgoing neighbors of an agent that share the focal agent's opinion will be greater than the number of neighbors that oppose the focal agent's opinion, because the rewiring dynamics transforms active links into inert ones. This implies that if an agent's opinion changes, the agent is likely to experience a subsequent loss of incoming links because the majority of neighbors now oppose the opinion and rewire their links with some probability. In the long run agents that frequently change their opinion have lower in-degree than those who change their opinion rarely. Therefore, the attention, measured in terms of incoming links, focuses on the agents that have a low out-degree and thus rarely change their opinion (\fref{fig:degrees}).

Focusing the attention on agents of low out-degree impedes the propagation of opinions across the network. In particular, it can lead to the formation of small clusters that have few outgoing links and hence have a very high resistance to invasion of the opposing opinion. In an extreme case small subgraphs can form in which all nodes are in consensus and all outgoing links starting within the subgraph lead to other nodes in the same subgraph. Such structures are attracting components of the network, which can be entered by following a directed path but never left due to the lack of outgoing links. Therefore, such subgraphs can never be invaded by the opposing opinion. Further, no outgoing links leaving the subgraph can be formed because none of the nodes in the subgraph will ever rewire an outgoing link.

We call subgraphs that are hard or impossible to invade self-stabilizing structures. The initial formation of such a structure through the creation of a state-homogeneous attracting component is a stochastic event that occurs with a small probability. However, once such a structure has been formed it can grow as other nodes rewire their outgoing links into the structure. Nodes of low out-degree can be recruited rapidly because only few rewiring events are necessary to rewire all of their outgoing links into the self-stabilizing structure. Recruitment of nodes with more outgoing links takes longer as more rewiring events are required. Note that the size of a self-stabilizing structure can only increase. In simulations, networks observed shortly before fragmentation are often found to consist of two almost disconnected clusters, which are only connected by a few nodes of high out-degree. Because of their frequent changes of state, these connecting hubs have very few or no incoming links.

\begin{figure}
 \includegraphics[width=0.48\textwidth]{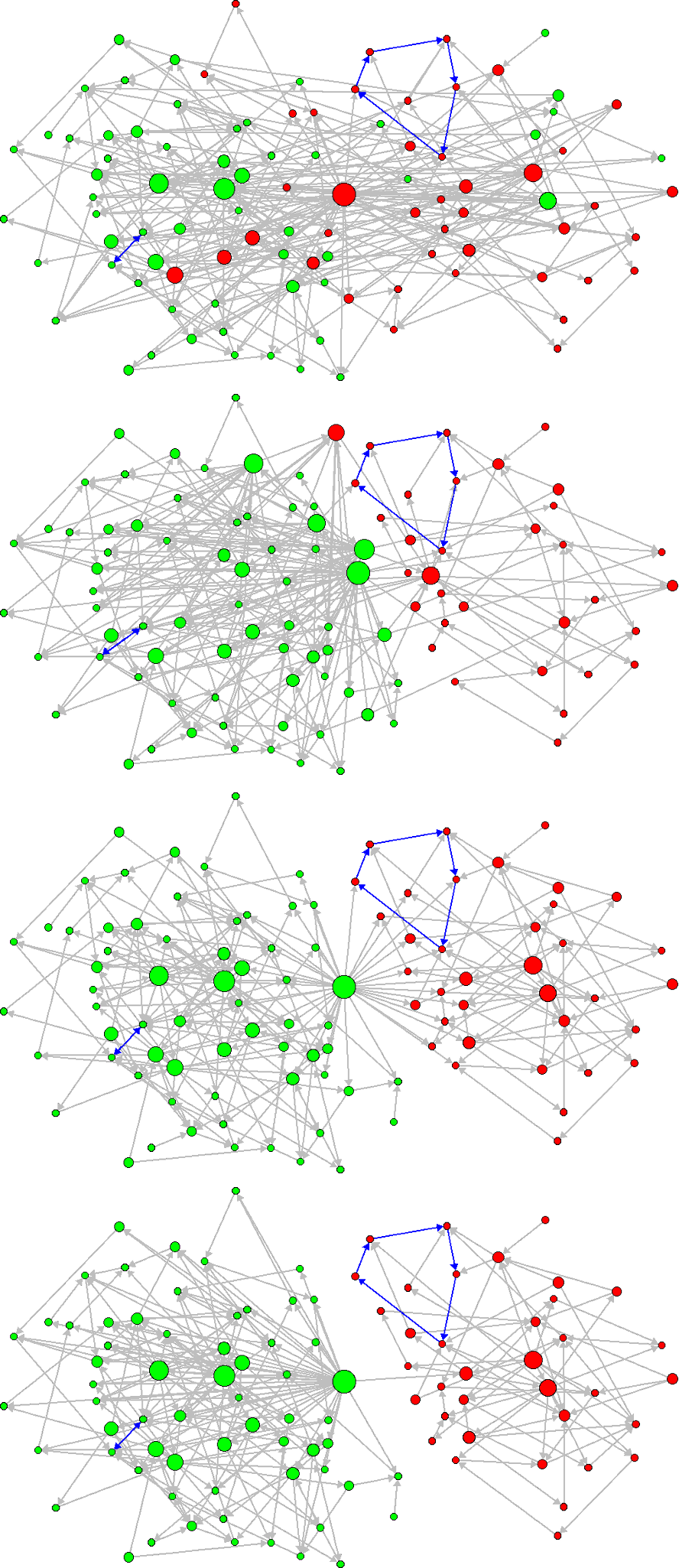}
 \caption{\label{fig:split-net-a2}(Color online) Snapshots of an exemplary network of $N=100$ nodes with out-degree distribution $P_{\rm out}(k)\propto k^{-2}$ at $t=200$, $250$, $350$, and $400$ (from top to bottom). Two state-homogeneous attracting components (blue/dark gray) nucleate self-stabilizing structures leading to early fragmentation. Note that the nodes with high out-degree (represented by the node size) have very low or zero in-degree and keep the network connected for a long time. Here, fragmentation is reached at $t\approx 413$.}
\end{figure}

For illustration of the mechanism described above an embedding of a small network at four time points before fragmentation is shown in \fref{fig:split-net-a2}. The network breaks into two almost disconnected clusters. The remaining connections are finally formed by a single hub node. The fragmentation has been nucleated by the formation of two state-homogeneous attracting components: a self-referential cycle consisting of five nodes of out-degree one and another consisting of two such nodes.

\begin{figure}
  \includegraphics[width=0.48\textwidth]{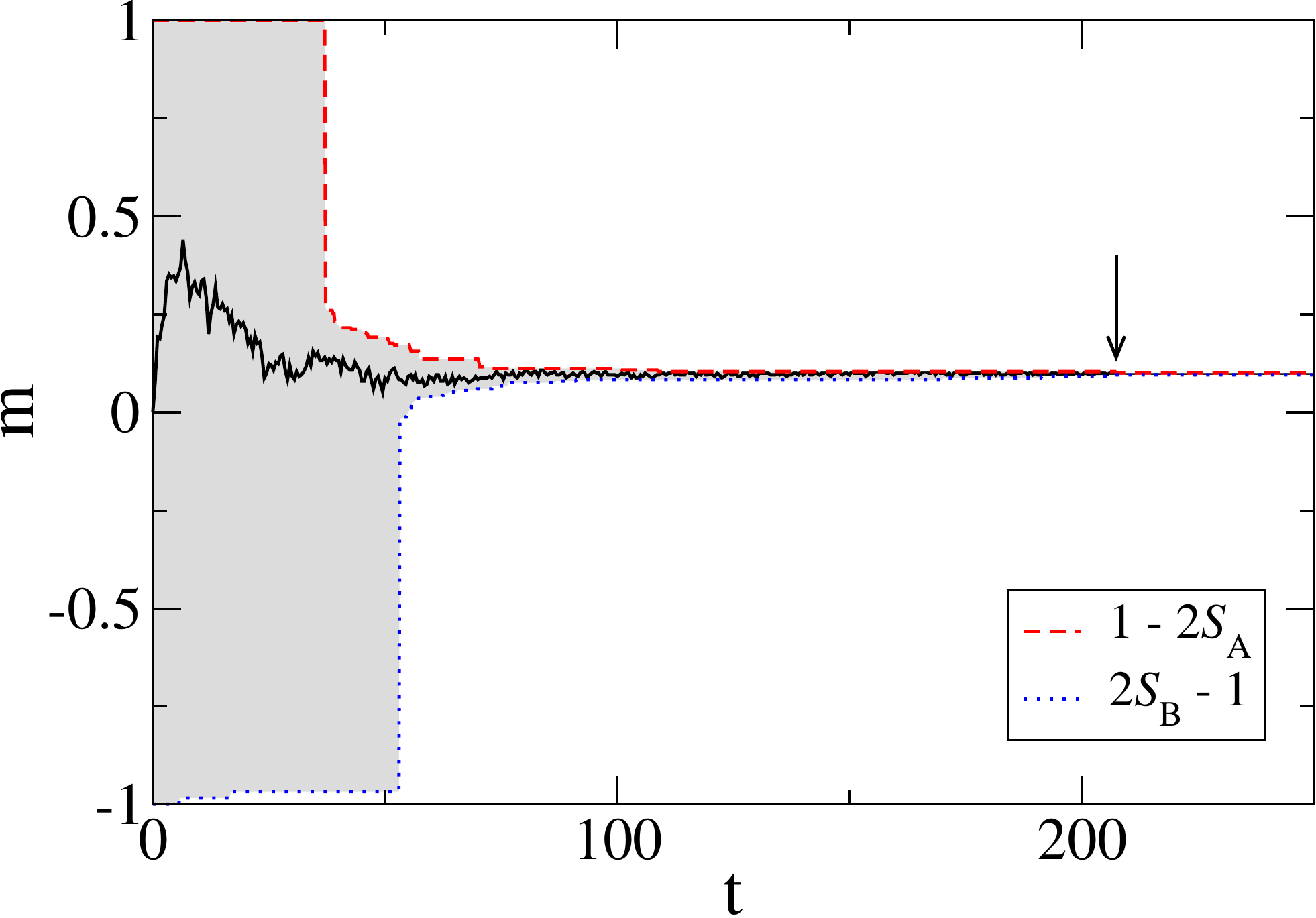}
  \caption{\label{fig:stable-components}(Color online) Typical restriction of the permissible magnetization range by self-stabilizing structures. Shown are the upper and lower bounds for $m$ (dashed red and dotted blue lines) computed from the total size $S_{\rm A}$ and $S_{\rm B}$ of the stable A- and B-components, i.e. maximally connected components in which each node has outgoing links to nodes in the same state only. The magnetization is confined to the shaded area and effectively arrested eventually at a fixed value. The arrow marks the point when only a single remaining hub node connects the two parts of the network (cf.\ \fref{fig:split-net-a2}). Here fragmentation is reached at $t\approx 524$. The jumps occur when the last links leaving a self-stabilizing structure are rewired to nodes within the same structure, so that it is fully stabilized and can be detected. The parameters are $N=500$, $P_{\rm out}(k)\propto k^{-2}$, and $p=0.3$.}
\end{figure}

Given the observations above we can explain the shape of the trajectories shown in \fref{fig:traj-lu}. Because the formation of a self-stabilizing structure is a rare event, they are generally not present in the initial network. The system therefore approaches the parabola of active states, which is in agreement with results from the undirected model and the analytical approximations for the directed model. However, while the system drifts along the parabola of active states, self-stabilizing structures are eventually formed due to the ongoing rewiring. As the self-stabilizing structures grow, the permissible range for the magnetization shrinks, effectively arresting $m$ as almost all nodes are recruited into the self-stabilizing structures (\fref{fig:stable-components}). Because the last nodes to join the structures are ``hub'' nodes with high out-degree, a relatively high density of active links can be maintained for some time. Because the hub nodes undergo rapid opinion switches, rewiring can only slowly separate them from opposing neighbors, which explains the slow fragmentation. The switching and rewiring of the hub nodes are clearly visible in the time series of the active link density $\rho$. In the inset in \fref{fig:traj-lu}, this is shown for the last $10^3$ time units before fragmentation. Here, one of the two remaining connecting hubs detaches from one of the components and the final hub still switches several times before eventually also separating.

Summarizing the observations above, we conjecture that early fragmentation is initiated by the formation of self-stabilizing structures among nodes of low out degree. We emphasize that contrary to most dynamical phenomena observed in scale-free networks, the dynamics of interest is generated primarily in the nodes of low degree. Nodes of high degree still play an important role as they are the last nodes to connect the separating components and thus determine the time of fragmentation. This mechanism is not captured by current analytical approaches, because it relies on the build-up of negative correlations between the in-degree and out-degree of nodes that is neglected in previously proposed approximation schemes.

\subsection{Poissonian out-degree distribution with low \texorpdfstring{$\boldsymbol\avk$}{<k>}}
Because the mechanism postulated above relies on the formation of correlations, one can perform a simple test by considering a system in which these correlations are removed by an additional rewiring process. However, such a test is difficult in scale-free networks for two reasons: First, because of the constraints in scale-free topology it is well-known that it is difficult to remove correlations in scale-free networks completely, and second, because of the presence of nodes of very high degree, fragmentation takes a long time, making numerical studies of fragmentation tedious.

Our reasoning above predicts that early fragmentation should be observed also in directed Poissonian networks with sufficiently low mean degree. In the present section we therefore consider a Poissonian network with a mean degree of 4, which avoids the difficulties encountered in scale-free networks. In this section we show (a) that this network exhibits early fragmentation and (b) that the early fragmentation can be avoided by an additional rewiring mechanism that destroys the correlations implicated in the formation of self-stabilizing structures.

\begin{figure}
 \includegraphics[width=0.48\textwidth]{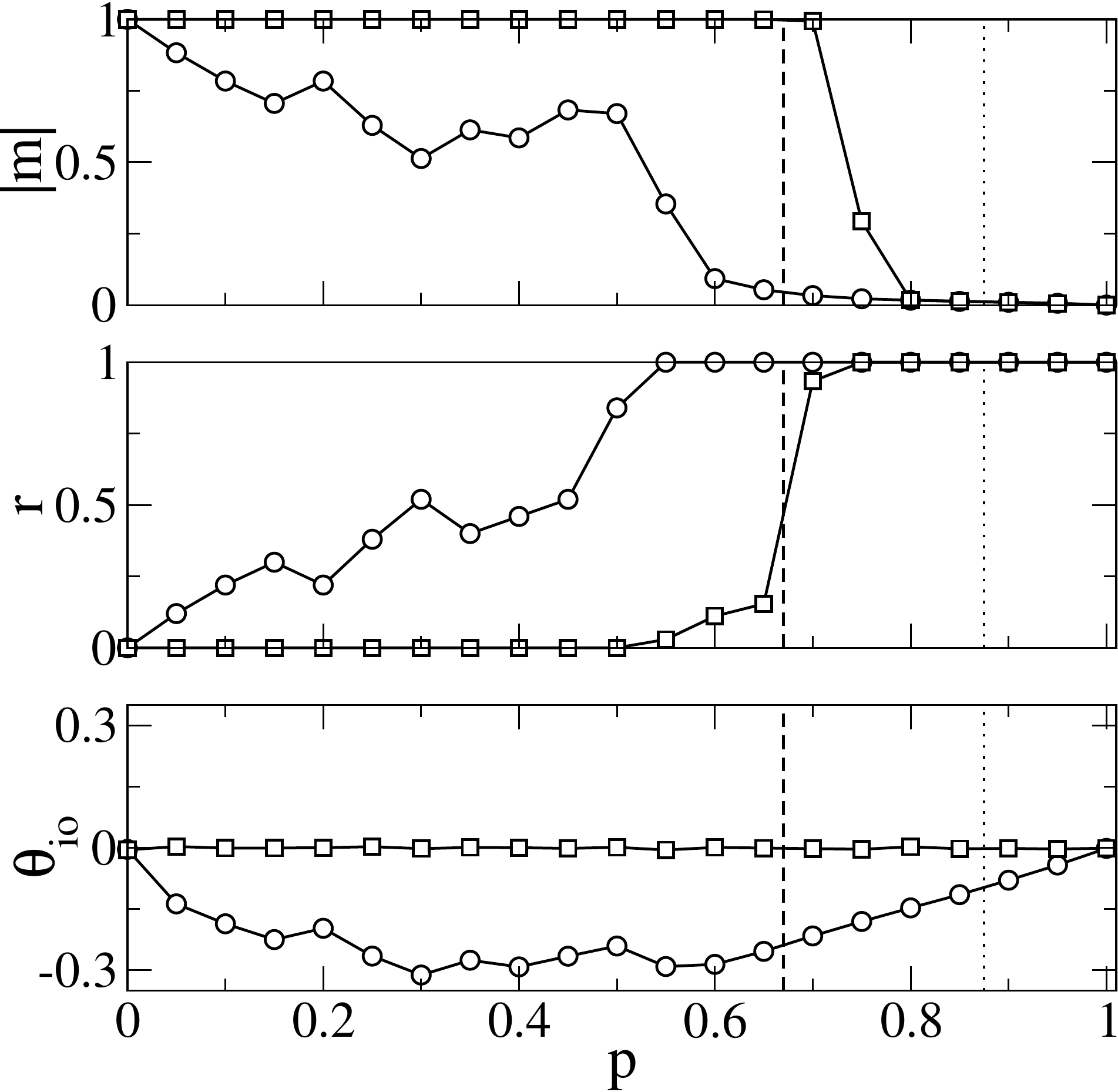}
 \caption{\label{fig:mag-mu4-inert} Fragmentation of networks with Poissonian out-degree distribution and low $\avk$, in which the inert links are also rewired (squares) or are not rewired (circles). Early fragmentation is clearly visible, although the shown averages over 100 simulation runs are still rather noisy due to the highly stochastic nature of early fragmentation. Shown are the absolute value of the magnetization (top), the proportion of fragmenting simulation runs (middle), and the correlation coefficient between the in- and out-degree of the nodes in the final state (bottom). The parameters are $N=10^4$, $\avk=4$, and $\hat k_{\rm in}= \hat k_{\rm out}=10$.}
\end{figure}

Simulation results for the network described above are shown in \fref{fig:mag-mu4-inert}. The figure shows clear evidence of fragmentation well below the estimated fragmentation point. Further, this early fragmentation is accompanied by the build-up of negative correlations between the in- and out- degrees of the nodes. This confirms our previous observation that attention focuses on those nodes who pay little attention to others themselves, as we have also seen in the scale-free case in \fref{fig:degrees}.

To verify that the correlation described above is the cause and not a symptom of the early fragmentation, we now consider a different variant of the model. This variant is identical to the model used so far, except that when an inert link is chosen, this link is also rewired to a randomly chosen target node that is in the same state as the source.

The model variant in which the rewiring of inert links is switched on shows no evidence for early fragmentation (see \fref{fig:mag-mu4-inert}). Fragmentation occurs in a relatively sharp transition at a critical rewiring rate $p_c$ that is consistent with the estimate from the motif expansion. We emphasize that the rewiring of inert links neither introduces nor destroys active links. It therefore has no \emph{direct} impact on fragmentation. However, rewiring inert links prevents the build-up of correlation between the in-degree and the out-degree of nodes and thereby inhibits the formation of self-stabilizing structures. The absence of early fragmentation in a model where these correlations are removed confirms the causal relationships postulated above. We therefore conclude that in directed adaptive networks the slow build-up of negative correlations between in-degree and out-degree can initiate early fragmentation by leading to the formation of self-stabilizing structures.

\section{Conclusions}
In the present paper, we have investigated an extension of the voter model on adaptive networks that takes the directionality of the interactions among the agents into account. We found that our model can transition to a fractionated state for rewiring rates that lie much below the critical value estimated using analytical approaches known to work well in undirected models.
We discovered that fragmentation occurs due to a novel mechanism that depends inherently on the directed nature of the links. This \emph{early fragmentation} occurs when agents focus their attention on those who are steady in their opinion because they pay attention only to few sources of information. In this case self-stabilizing structures can form that nucleate fragmentation.

Our results illustrate that directed networks can exhibit new physics not observed in their undirected counterparts. Especially in the investigation of opinion formation processes, the often directed flow of attention should therefore be taken into account in models.

In the context of real-world opinion formation processes the mechanism described here may constitute a threat but also an opportunity. Early fragmentation maintains diversity of opinions, it may thus aid the survival of counter-factual myths, but also of legitimate and well-founded views of minorities. It is conceivable that this mechanism may be employed in the future to adjust the perceived degree of controversy in controlled environments. For instance, in online discussion boards or music recommendation systems the underlying software can in principle control which posts are displayed to whom. It could thus encourage rewiring of attention that either facilitates or inhibits early fragmentation.

Even the adaptive directed voter model paints a highly simplified picture of real-world opinion formation processes and thus must be considered as a toy model. Therefore, investigation of more realistic models is an important goal for the future. Based on the results and analysis presented in the present paper we believe that the mechanism of early fragmentation will be observed whenever directed attention is focused preferentially on agents that change their opinions at less than average rate. We therefore expect that early fragmentation should be robust to future refinements of the model.

A key ingredient that is missing in our present model is novelty. Here we considered only the exchange of opinions regarding a single well defined question, whereas in reality many discussions are enriched by the constant inflow of new ideas. We have shown that homophily favors connecting to poorly informed agents and thereby promotes early fragmentation, whereas curiosity would favor connecting to well informed agents and thereby hinder early fragmentation. In this light, novelty, whether in the form of true innovation or arbitrarily changing fashions may play an important role in preventing social fragmentation.

The authors thank G.\ Demirel for fruitful discussions. The work of CH was partially supported by the National Science Foundation under Grant No.\ PHY-0848755.

%

\end{document}